\documentclass[conference]{IEEEtran}
\IEEEoverridecommandlockouts
\usepackage{dcolumn}
\usepackage{graphicx}
\usepackage{siunitx}
\usepackage{balance}
\usepackage{multirow}
\usepackage[font={small},labelfont=small]{caption}

\ifCLASSINFOpdf
\else
\fi

\usepackage{multirow}

\usepackage{amsmath}
\usepackage{algorithmic}

\usepackage{array}
\usepackage{textcase}
\usepackage[tablename=TABLE]{caption}
\DeclareCaptionTextFormat{up}{\MakeTextUppercase{#1}}
\usepackage{comment}
\usepackage{tabularx,booktabs}
\usepackage[left=0.625in,right=0.625in,bottom=0.90in,top=0.9in]{geometry}
\begin{document}
\bstctlcite{IEEEexample:BSTcontrol}
%
\title{Revisiting Data Recovery Loops in 6G Networks}
%
%
%


\author{

\IEEEauthorblockN{Uyoata E. Uyoata\IEEEauthorrefmark{1}\IEEEauthorrefmark{3} Abolfazl Amiri\IEEEauthorrefmark{2}\IEEEauthorrefmark{4}, Enric Juan\IEEEauthorrefmark{2}, Guillermo Pocovi\IEEEauthorrefmark{2}, }
\IEEEauthorblockN{Pilar Andres-Maldonado\IEEEauthorrefmark{2}, Klaus I. Pedersen\IEEEauthorrefmark{1}\IEEEauthorrefmark{2}, and Troels  Kolding\IEEEauthorrefmark{2}}

\IEEEauthorblockA{\IEEEauthorrefmark{1}Wireless Communication Networks Section, Department of Electronic Systems, Aalborg University, Denmark}

\IEEEauthorblockA{\IEEEauthorrefmark{2}Nokia, Aalborg, Denmark}

\IEEEauthorblockA{\IEEEauthorrefmark{3}ueu@es.aau.dk}
\IEEEauthorblockA{\IEEEauthorrefmark{4}abolfazl.amiri@nokia.com}

}

\markboth{Journal of \LaTeX\ Class Files,~Vol.~14, No.~8, August~2015}%
{Shell \MakeLowercase{\textit{et al.}}: Bare Demo of IEEEtran.cls for IEEE Journals}
%



\maketitle

\begin{abstract}
Mechanisms for data recovery and packet reliability are essential components of the upcoming 6$^\text{th}$ generation (6G) communication system.
In this paper, we evaluate the interaction between a fast hybrid automatic repeat request (HARQ) scheme, present in the physical and medium access control layers, and a higher layer automatic repeat request (ARQ) scheme which may be present in the radio link control layer. 
Through extensive system-level simulations, we show that despite its higher complexity, a fast HARQ scheme yields $>\SI{66}{\percent}$ downlink average user throughput gains over simpler solutions without energy combining gains and orders of magnitude larger gains for users in challenging radio conditions. We present results for the design trade-off between HARQ and higher-layer data recovery mechanisms in the presence of realistic control and data channel errors, network delays, and transport protocols. We derive that, with a suitable design of 6G control and data channels reaching residual errors at the medium access control layer of $5 \times 10^{-5}$ or better, a higher layer data recovery mechanism can be disabled. We then derive design targets for 6G control channel design, as well as promising enhancements to 6G higher layer data recovery to extend support for latency-intolerant services.

\end{abstract}


\IEEEpeerreviewmaketitle

\section{Introduction}
Data recovery, i.e., the ability of a communications system to recover lost packets or transmissions, is particularly important in wireless networks where the channel quality can vary quickly and connectivity can be intermittent for example at the cell edge. One of the most successful data recovery concepts is the hybrid automatic repeat request (HARQ) mechanism \cite{9471818}. HARQ allows the network to quickly identify a transmission error and to perform a fast retransmission of a failed packet allowing the receiving end to combine several transmissions to incrementally improve the probability of successfully receiving said packet. As such, HARQ combines forward error correction (FEC) with automatic repeat request (ARQ) mechanisms. For cellular networks, HARQ was introduced as part of high-speed packet access in 3G in \num{2012} and has been a key component in 4$^\text{th}$ generation (4G) and 5$^\text{th}$ generation (5G) system designs. HARQ has evolved significantly towards supporting \num{5} and \num{6} nines of reliability while supporting millisecond level packet latency, for example, the ultra-reliable low latency communications (URLLC) mode in 5G \cite{8010757}.

 \begin{figure}[ht]
 \centering
 \includegraphics[width=\linewidth,trim={4.5cm 2cm 7cm 1cm},clip]{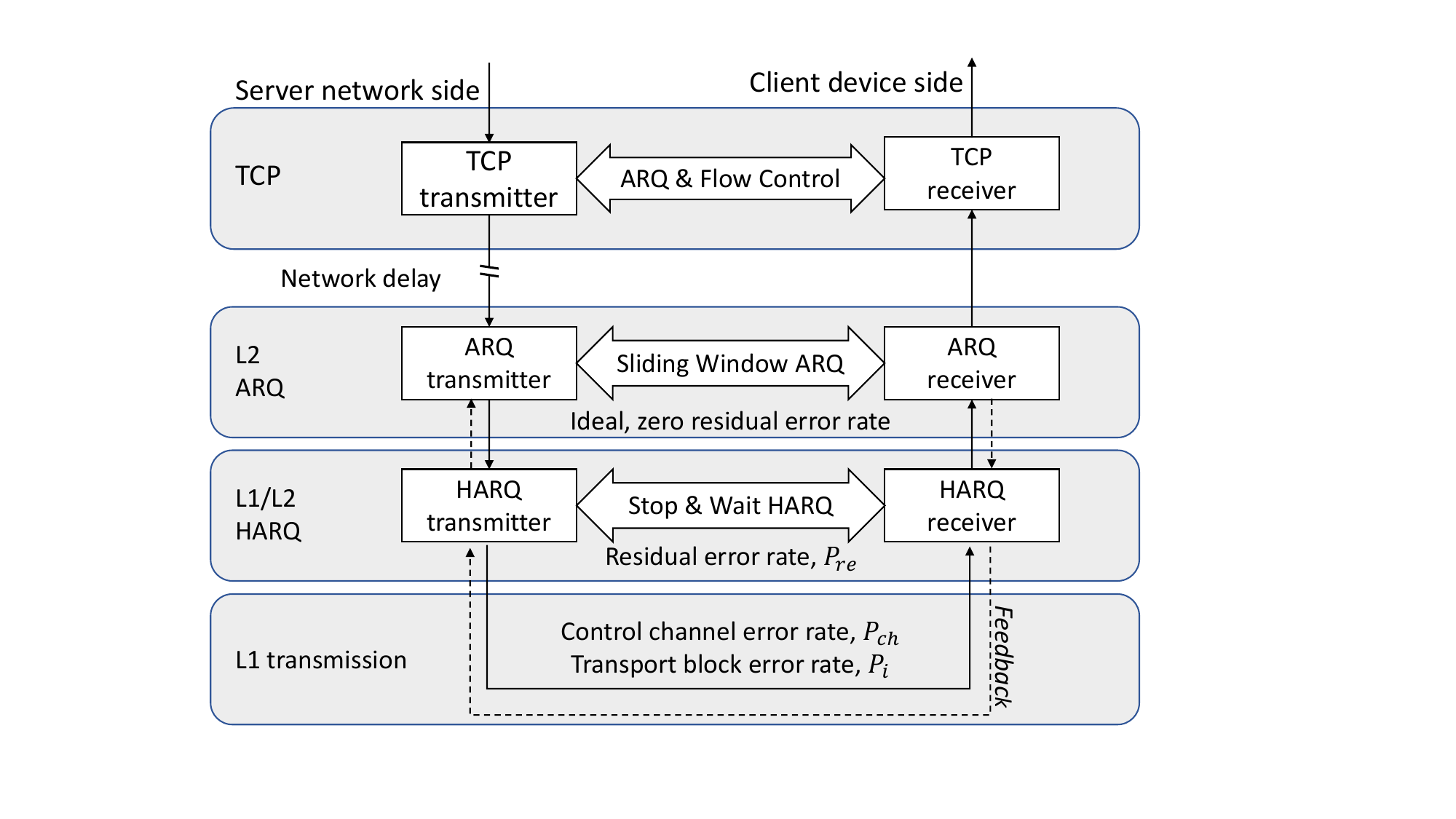}
 \caption{HARQ, ARQ, and TCP Data Recovery Loops}
 \label{fig_00}
 \end{figure}

Although HARQ is very capable, there will typically be few residual errors e.g., due to control channel imperfections. Removing these errors is not possible in a given deployment or would lead to undesirable trade-offs in terms of capacity or delays. One example of a residual error is when the base station or the next generation node-B (gNB) in 5G, misinterprets the feedback signal from the user equipment (UE) side as a positive acknowledgment (ACK) when the UE had sent a negative acknowledgment (NACK) indicating that transmission was not correctly decoded \cite{8347007}. 

Therefore, wireless networks offer a second layer of data recovery that monitors errors by tracking packet sequence numbers and initiates retransmissions separate from HARQ. In 5G, such recovery is handled at the radio link control (RLC) layer, using ARQ based on a sliding window technique, when RLC is operated in the acknowledged mode (AM) \cite{3gpp38322}. When RLC is operated in the unacknowledged mode (UM), there is no ARQ. While such processes often induce significant packet delays and higher transmission overhead, when compared to the HARQ loop, they offer independent error handling to achieve or guarantee near-flawless transmissions. The hierarchy of HARQ, ARQ, and TCP data recovery loops is shown in Fig. \ref{fig_00}, and the various details regarding feedback mechanisms and residual errors are further explained throughout this paper.

In 6G, data recovery mechanisms should be revisited taking into account new requirements, available enhancements to the physical layer, as well as the ongoing evolution of transport-level protocols. Main new use cases in 6G are envisioned to have heightened sensitivity to latency, rendering existing higher-layer ARQ methods ineffective. Consequently, a decision must be made to either disable or redesign ARQ. Transport level protocols, such as more recent variants of TCP, have evolved to be more robust to packet losses and re-ordering errors \cite{9187252}.

Both HARQ and ARQ, as well as their combination, have been researched in the past. In \cite{7600947} interactions between RLC modes and transport protocols (UDP and TCP) are studied for a 4G network. Results show that for ideal control channels, i.e. the physical control format indicator channel and physical downlink control channel (PDCCH), the combination of UDP and RLC acknowledged mode (AM) offers nearly the same throughput as running the network with RLC unacknowledged mode (UM) with the TCP transport protocol. Moreover, when the PDCCH is erroneous, the average throughput significantly drops for the combination of TCP with RLC UM owing to the reduction of the congestion window due to unrecovered packets at the lower layer RLC. The work in \cite{7600947} does not detail the assumed HARQ configurations, such as the target block error rate, which makes the detailed interaction between HARQ and RLC with TCP unclear. 

In \cite{8116400}, the interaction between HARQ and RLC data recovery loops is investigated when TCP CUBIC congestion control algorithm is employed for millimeter wave links. The results show that the spectral efficiency improvement from RLC ARQ on top of HARQ is significant at farther distances from the base station. This throughput improvement is more due to HARQ than ARQ since there is only a slight performance difference between having ARQ and operating in the RLC UM. Furthermore, the results indicate that for a UE connected to both a 4G network at a lower carrier frequency and a mmWave small cell, a lower carrier cell provides a more reliable path for TCP-based transmissions. In \cite{7562173}, it is shown that lower layer data recovery loops like HARQ and ARQ significantly reduce errors that higher layer mechanisms like TCP need to handle, even in mmWave scenarios that are characterized by channel variations arising from signal blocking. The performance of some TCP variants in 5G mmWave is evaluated for various blockage configurations in an urban environment in \cite{10.1145/3416011.3424749} where simulation results show that user mobility can affect TCP throughput in non-line of sight state. Such effect of user mobility is also shown in \cite{tcp_body} for a case of human body blockage.

In this paper, we present conclusions from detailed end-to-end and 3$^\text{rd}$ generation partnership project (3GPP) model-compliant system-level simulation results. A detailed model for residual error rate after HARQ as a
combination of uncertainties in both uplink and downlink control channels is presented.
The paper identifies the residual error rate boundaries where second-level data recovery will add only negligible benefit and thus can be disabled for applications where there is no latency budget to support its use. From these results, we derive reliability design targets for 6G control channel design and signaling procedures related to data recovery.  Vice versa, we show how much the ARQ mechanism can improve throughput for deployments of scenarios where tight reliability targets cannot be met. 
To the best of our knowledge, this is the first work to consider explicitly the interactions of the recovery loops given 6G networks with non-ideal physical uplink and downlink control channels using a dynamic system-level simulator fully compliant with 3GPP evaluation methodologies. Due to the dynamic nature of the system considered in this paper, exact closed-form theoretical solutions are not obtainable, hence simulations are used for evaluating results. 

The paper is organized as follows: In Section II, the data recovery loops considered in the research are explained. The system model used for results generation is described in Section III and results with analysis are given in Section IV. Design recommendations towards 6G are given in Section V before we conclude the paper in Section VI.

\section{Data Recovery Loops}
\subsection{Hybrid Automatic Repeat reQuest}

HARQ, present in the physical and medium access control (PHY and MAC) layers, enables fast data recovery using FEC and ARQ. Retransmissions of a transmitted transport block (TB) (or a code block group) that was not successfully decoded are requested through explicit feedback to the transmitter. The transmitter can send copies of the transmitted data or versions of the initially transmitted data with varied levels of redundancy which are combined at the receiver. There are variants of HARQ combining schemes including Chase combining and incremental redundancy (IR) \cite{Frede2002}. Retransmission happens until either the transmitting side receives a positive acknowledgment or the maximum number of retransmissions is reached if configured. For cellular communications, HARQ uses parallel processes, known as stop-and-wait channels, to allow for processing and response time without stalling data transmissions. With each retransmission, the chances of error in the combined transport blocks reduce, depending on the scheme and channel conditions. For effects and modeling approaches, see \cite{Frede2002}. The gains of retransmission come at a cost of overhead from multiple transmissions and control signaling as well as latency. For some services, such as extended reality (XR) and URLLC in 5G, the maximum number of retransmissions is limited to e.g. one or two before the latency budget is exceeded. 

There are some error cases where the HARQ loop is unable to correct transmission failures which lead to residual errors. These errors originate mainly from errors related to non-ideal control channels. One case is where the UE misses a (re)transmission by not detecting properly a resource grant provided on the downlink control channel, i.e. the PDCCH in 5G, and where the UE therefore does not send a NACK on the feedback channel. This non-transmission, or discontinuous transmission (DTX) related to the gNB expectations, may still be interpreted by an ACK by the transmitter, the so-called DTX to ACK error. When the gNB interprets DTX as an ACK, it skips sending the retransmission grant and so results in a missing TB. 
A second case of residual error is when the UE sends back a NACK but the transmitter interprets it as an ACK, i.e. the so-called NACK to ACK error. 
The probability of DTX to ACK and NACK to ACK errors are typically between \SI{1}{\%} and \SI{0.01}{\%}, respectively, while the probability of missing a resource grant is typically in the 1\%\ range \cite{7063628}. 

The probability of residual error after HARQ can be expressed as;
\begin{equation}
P_{re}\! = \!\!(1 - P_{ch})P_{e}P_{na} + P_{ch}P_{da} + P_{gu} + (1 - P_{ch})(1-P_{e})P_{an},
\label{eq_1}
\end{equation}
where $P_{ch}$ is the probability of error in the control channel carrying the resource grant (i.e. in 5G the PDCCH), and $P_{e}$ is the average probability of error of transmission over the air interface, in practice close to the initially configured block error rate (BLER). $P_{na}$ is the NACK to ACK error rate while $P_{an}$ models the reverse case of ACK to NACK error rate, and $P_{da}$ is the DTX to ACK error probability (see 
Fig.~\ref{fig_00} for an illustration of the non-ideal recovery loops).  
All the terms in equation (\ref{eq_1}) except $P_{ch}P_{da}$ capture the case where the downlink resource grant is correctly decoded by the UE whereas the remaining term is for the case where the resource grant is not decoded correctly and the feedback to request a re-sending of the resource grant is mis-detected. Depending on the implementation, residual errors may also arise if the initial transmission was so far off the actual channel conditions that recovery was not possible after a maximum amount of retransmissions configured by the system. Hence, the gNB gives up re-transmitting a certain transport block (TB) when a pre-configured maximum allowed number of retransmissions is reached. The probability of this event is given by $P_{gu}$. If the initial  BLER is set at $P_{e}$, then $P_{gu}$ after $n$ retransmissions is given by:
\begin{equation}
P_{gu}=(P_{e})^n,
\label{eq_2}
\end{equation}
For $P_{e} =$ \SI{0.1} and $n = 6$ for example, successful transmission is highly likely making  $P_{gu}$ negligible. Moreover, the impact of interpreting a successful delivery as a failed one ($P_{an}$) is not realized harmfully on the recovery loops and does not influence the BLER. Hence, equation \ref{eq_1} can be approximated with the most dominant and crucial terms as   
\begin{equation}
P_{re} \simeq (1 - P_{ch})P_{e}P_{na} + P_{ch}P_{da}.
\label{eq_3}
\end{equation}


\subsection{Radio Link Control Automatic Repeat reQuest (ARQ)}
To handle residual errors from the HARQ, a higher layer recovery loop may be used. In 5G, ARQ is implemented at the RLC layer which, besides basic functions of segmentation and reassembly, takes care of ARQ when operated in the AM \cite{3gpp38322}. Retransmissions in the MAC layer differ from RLC retransmissions in that RLC retransmissions are slower and are not triggered per lost transport block, but rather based on a sliding window mechanism implemented by timers and events. Status protocol data units (PDUs) are sent by the receiver to the transmitter either at the expiry of a set reassembly timer or at the reception of a polling request from the transmitter. The reassembly timer is initialized on the detection of a missing PDU which is possible with the help of assigned sequence numbers and segment indicators. There is no combining gain available with this solution. The increased reliability that RLC ARQ guarantees comes at the cost of latency, typically tens of milliseconds in 5G. In principle, there can be residual error cases also from RLC ARQ, but they can be designed to be virtually zero and are ignored in this study. 

\subsection{Transport Control Protocol}
With cellular networks offering primarily mobile broadband, TCP is used for the majority of traffic, i.e. more than 90\%\ as shown in \cite{Schumann2022}.
TCP congestion control algorithms treat packet loss (from non-receipt of acknowledgment) as congestion and thus reduce the number of transmitted packets in reaction to the perceived congestion. Flow control is also a feature of TCP that helps regulate the amount of packets transmitted concerning the buffer capacity of the receiver. In this study, we compare the following TCP flavors:
\subsubsection{TCP Reno} This is the baseline congestion control algorithm built for low-speed wired networks and adapts the congestion window linearly after exiting a slow start phase. The congestion window is initialized to a small value that is increased per received acknowledgment. The trigger for reducing the congestion window (by half) is either the non-receipt of an acknowledgment within a retransmission timeout or the receipt of duplicate acknowledgments.  The additive approach employed by TCP Reno is less implementation intensive but falls short of the demands of high bandwidth networks. This paper uses the IETF RFC 5681 version of TCP Reno as a baseline \cite{rfc5681}. 
\subsubsection{TCP CUBIC} This congestion control algorithm increases the congestion window using a cubic function instead of the linear function that its predecessors employed. By alternately increasing and decreasing the aggressiveness of its window growth depending on its distance from the point of last congestion, TCP CUBIC can offer scalability and fairness improvements for fast networks \cite{Ha2008CUBICAN}.

\section{System Model}
Extensive simulations were carried out using a 3GPP-compliant system-level simulator \cite{SLS_survey}. The main simulation assumptions are given in Table \ref{Tsimo}.  
The simulator models the cellular communication protocol stack with end-to-end packet flows from when packets arrive at the packet data convergence protocol (PDCP) layer through to the physical layer. 
The simulator operates with OFDM symbol time resolution and it has been used in previous scientific papers (e.g., \cite{9732349}) and for 3GPP standardization activities. 

These simulations are conducted such that the same simulation is repeated \num{10} times using different pseudo-random seeds. After that, the simulation results are combined to ensure sufficient statistical confidence.

The simulated network environment is a dense urban macro environment with \SI{200}{m} inter-site distance having \num{7} base stations each with three sectors making a total of \num{21} cells, serving \num{21} uniformly distributed UEs moving at a speed of \SI{3}{km/h}. We are interested in deployments with high available capacity to clearly see performance degradation for high-rate TCP flows. Antenna configurations at the base stations are based on 3GPP TR36.814 in \cite{3gpp36814} and the maximum transmit power of each base station is fixed at \SI{51}{dBm} to ensure congruence with typical 5G assumptions. Packets are scheduled in the frequency domain using the proportional fair scheduler. We use the FTP Model 3 traffic generator which is characterized by generating packets of a given size and with a Poisson inter-arrival rate, typically assumed for internet and IP traffic. Note that our choice of file size of  \SI{35}{MB} with an arrival rate of \num{0.25} packets per second results in an offered load of \SI{70}{Mbps} per user, which is premised on remaining in the feasible load region of the network. 


\section{Performance Analysis}
In this section, the results from extensive system-level simulations are presented and discussed. The dynamic nature of the considered end-to-end system makes exact closed-form solutions impracticable, therefore simulations are used. Key performance indicators include average throughput per user and FTP3 packet and are described as follows,
\begin{itemize}
    \item User throughput: is the average application layer throughput (goodput) for a user measured over the time from when the user enters the system until it leaves.
    \item  Throughput per FTP packet: is the application layer throughput for transmission of one FTP3 packet and the average is calculated among all the transmitted packets of all the users in the system.
\end{itemize}

\begin{table}[t]
\caption{Simulation Parameters} 
\centering 
\begin{tabular}{l||l} 
\hline\hline 
Parameter & Value\\ [0.5ex]
\hline 
Network Environment
 & 3GPP Dense Urban  with \num{21} \\
  & macro cells
\\ 
BS Configuration
 & \num{7} sites with \num{3} sectors per site \\

 & and 200 m ISD, \\
 & \SI{51}{dBm} Tx power, \num{32} TX and
\\
 &  Rx antennas
\\
Carrier configuration
& \SI{4}{GHz} carrier frequency with \\

&  \SI{100}{MHz}\\
& bandwidth and \SI{30}{\kHz} Subcarrier \\
&  Spacing\\
Cell loading &Even load of 1 UE per cell\\
\hline
Data Channel & \num{0.1} BLER \\ 
& target for first transmissions\\
\hline
Control Channel & NACK to ACK Error: \SIrange[range-phrase=-]{0}{1}{}\\

\hline
Antenna Configuration
 & Single-user MIMO (rank 1) with\\
& LMMSE-IRC receiver at the UE\\
 \hline
Packet Scheduler
 & Proportional Fair in\\

 & the frequency domain\\
Link Adaptation & Based on UE sub-band  \\
 & CQI reports every 5 ms,  \\
& and  Outer loop link adaptation \\
&(OLLA) based on UE ACK/NACK \\
&feedback\\
HARQ & Max. \num{6} HARQ Retransmissions \\
&and 8 HARQ processes\\
& Async HARQ with CHASE  \\
& (Gain = \num{0.95}) \\
RLC-ARQ& Max. ReTX = \num{13}, \\
& Poll Retransmit Timer: \SI{25}{ms}\\
& Reassembly timer: \SI{50}{ms}\\
\hline
UE configuration & Uniformly distributed \num{21} UEs;\\
& \SI{3}{km/h} UE speed\\
& \num{4} Rx antennas, \num{2} Tx Antennas\\

\hline
Traffic Model& \num{35} MBytes FTP3 file, \\
& \num{0.25} packets per second\\
\hline
Transport Model& Maximum segment size = \num{1500} Bytes, \\
&TCP Reno (Initial congestion \\
&window size = \num{3} MSS) \\
&CUBIC ($\beta=0.2$, $C=0.4$), \\
&Slow Start Threshold: \num{500} MSS, \\
&Network Delay: \SI{10}{ms}\\
\hline
Simulation Time & \num{60} seconds, Number of Drops: \num{10},\\
 &  \num{4} seconds warm-up time\\
\hline 
\end{tabular}
\label{Tsimo}
\end{table}

\subsection{Results Discussion}
\begin{figure}[!t]
\centering
\includegraphics[width=\linewidth]{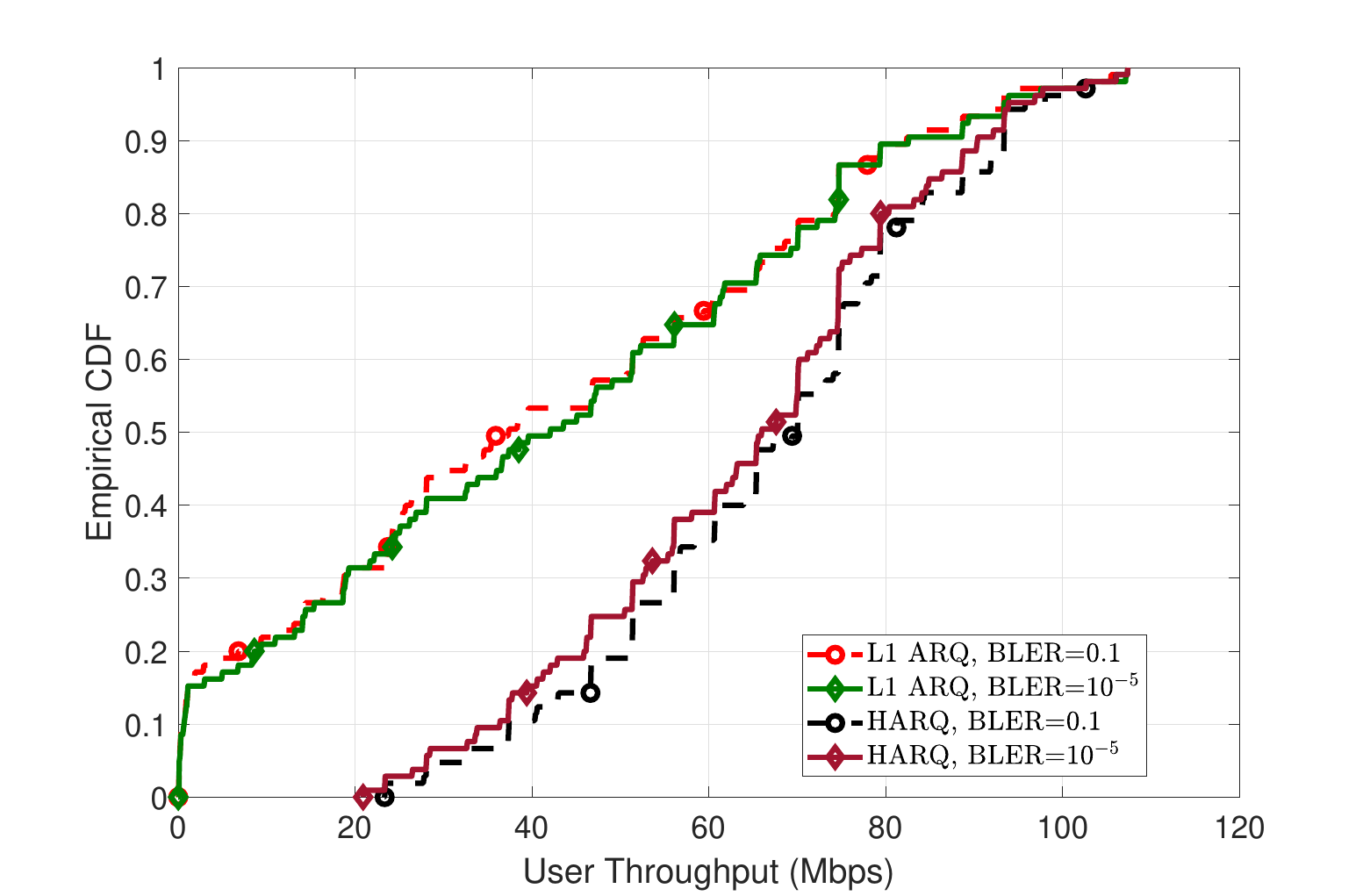}
\caption{Empirical CDF of user throughput for L1 ARQ and HARQ methods with different configured BLER targets ($0.1$ and $10^{-5}$).}
\label{fig_harq_l1arq}
\end{figure}

 To understand the importance of HARQ in good network design, Fig. \ref{fig_harq_l1arq} compares the empirical cumulative distribution function (CDF) of the user throughput of an HARQ-enabled setup with a setup where only ARQ is implemented at layer 1 (with ideal detection of each transmission error and immediate retransmission for each failed TB). In this layer 1 ARQ, there is no combining of retransmissions, and as such, retransmissions are undesirable from the spectral efficiency perspective. We assume for the ARQ mode that the channel state information (CSI) measuring report from the UE guides what TB size can be supported based on an estimated block error rate (BLER) target of \num{0.1}. An outer loop link adaptation target (providing a dynamic offset to user CSI reports) is also configured to a BLER of $\num{1e-5}$ and $\num{0.1}$, respectively.
 
 As seen in the figure, the $50^{th}$ percentile results show $\num{66}\%$ improvement in user throughput utilizing combining gain being available from HARQ. It is also clear that for UEs that are in more challenging conditions, i.e. near the cell edge, represented by the $5^{th}$ percentile throughput numbers, ARQ at layer 1 is not capable of delivering any appreciable throughput. Simply replacing HARQ with even an ideal layer 1 ARQ scheme leads to unacceptable performance loss, and thus, the 6G baseline should allow for signal combining via HARQ.

\begin{figure}[!t]
\centering
 \includegraphics[width=\linewidth]{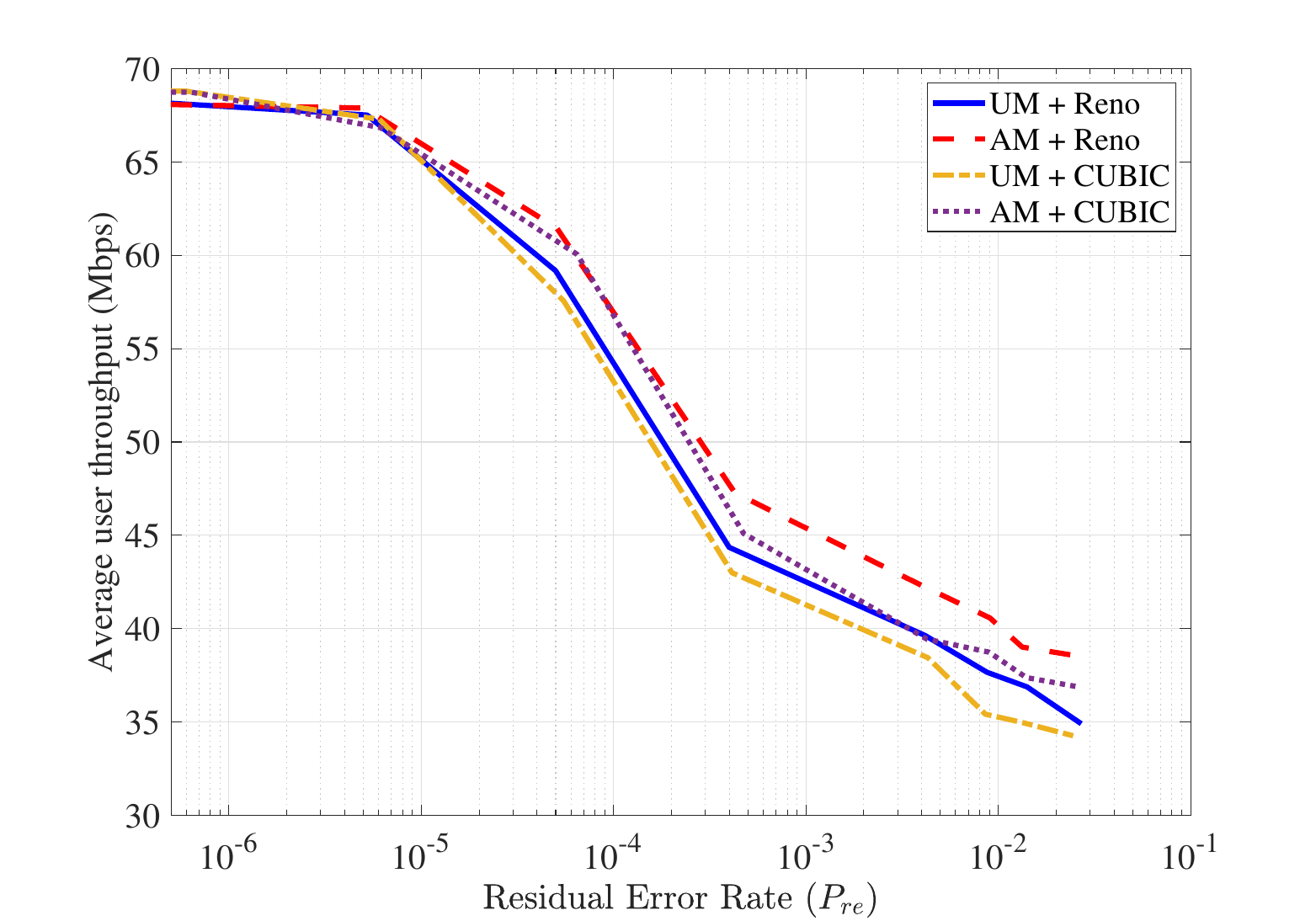}
\caption{Average user throughput (Mbps) versus HARQ residual error rate}
\label{fig_sim1}
\end{figure}

Fig.~\ref{fig_sim1} plots the relation between average user throughput versus HARQ residual errors for the two variants of TCP (Reno, Cubic) and the two modes of RLC (AM, UM). Different values for residual error rate are accomplished by varying the NACK to ACK error rate in the simulator while keeping the downlink control channel error rate fixed. The trend of the curves indicates a downward trajectory of user throughput as the residual error rate is increased. 
The ability of the network to reach the offered load or capacity (of \SI{70}{Mbps} as configured by the chosen traffic model) is reduced as the residual error is increased. The curves show the value of ARQ as the residual error rate is increased. Gain is more noticeable from 
$5\times10^{-5}$  HARQ residual error rate, where RLC ARQ offers \SI{5}{\%} or better improvement in throughput for each TCP variant. The reason for the improvement is that RLC ARQ can recover missing packets resulting from false positive events associated with HARQ, thus the gains become significant as the residual error rate increases.

\begin{figure}[!t]
\centering
\includegraphics[width=\linewidth]{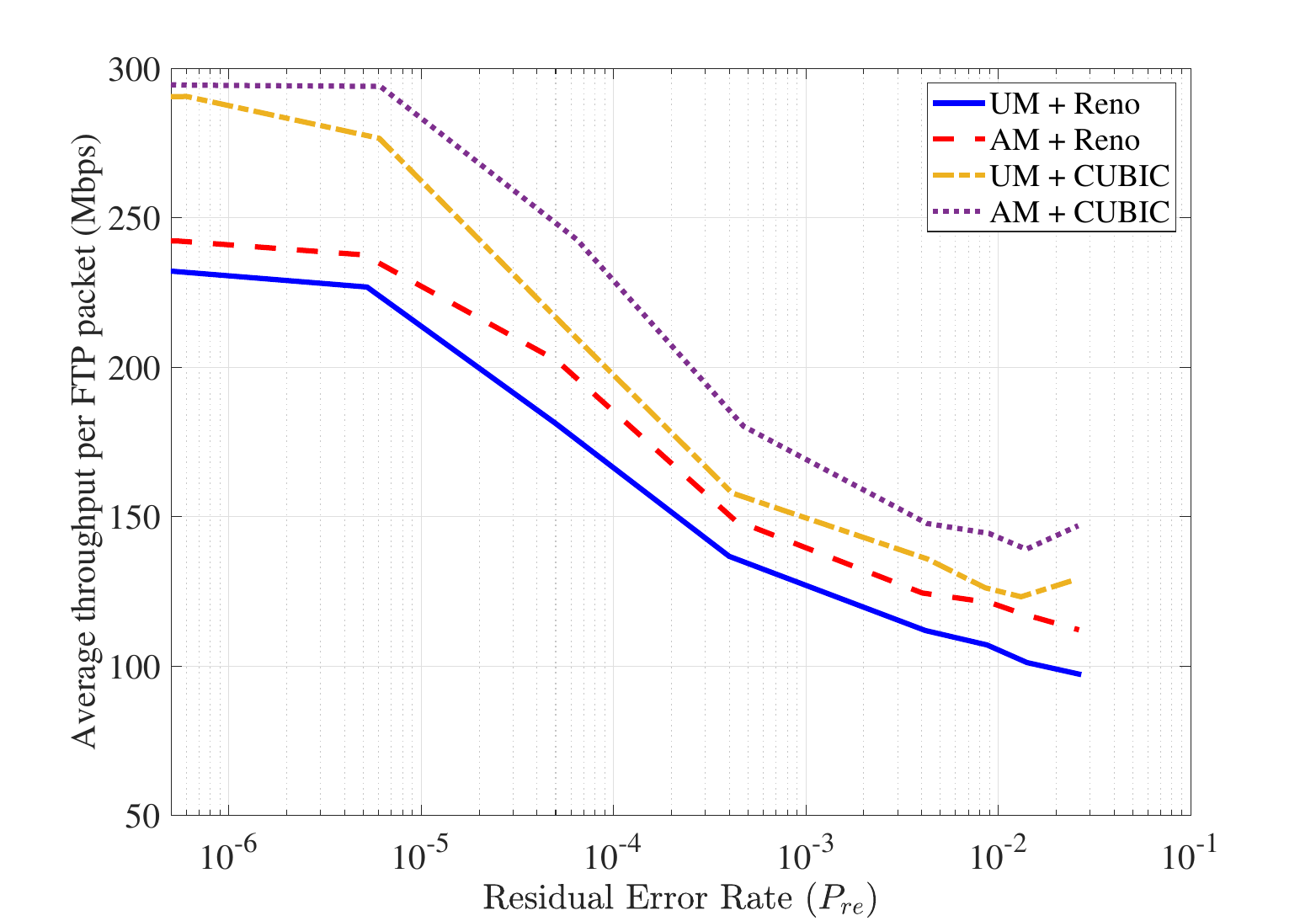}
\caption{Average throughput per FTP packet versus HARQ residual error rate}
\label{fig_sim3}
\end{figure}

Fig.~\ref{fig_sim3} shows the average throughput per FTP packet transmission in Mbps. Similar to Fig.~\ref{fig_sim1},  an increased HARQ residual error rate reduces the throughput for all combinations of transport protocol variants and RLC modes. As shown, the Cubic variant manages to transfer each packet with higher data rates than the Reno scheme. Due to this, the UEs spend more time idle after the packet transmission is done in the Cubic mode which results in a lower average user throughput illustrated in Fig.~\ref{fig_sim1}.

\begin{figure}[!t]
\centering
\includegraphics[width=\linewidth]{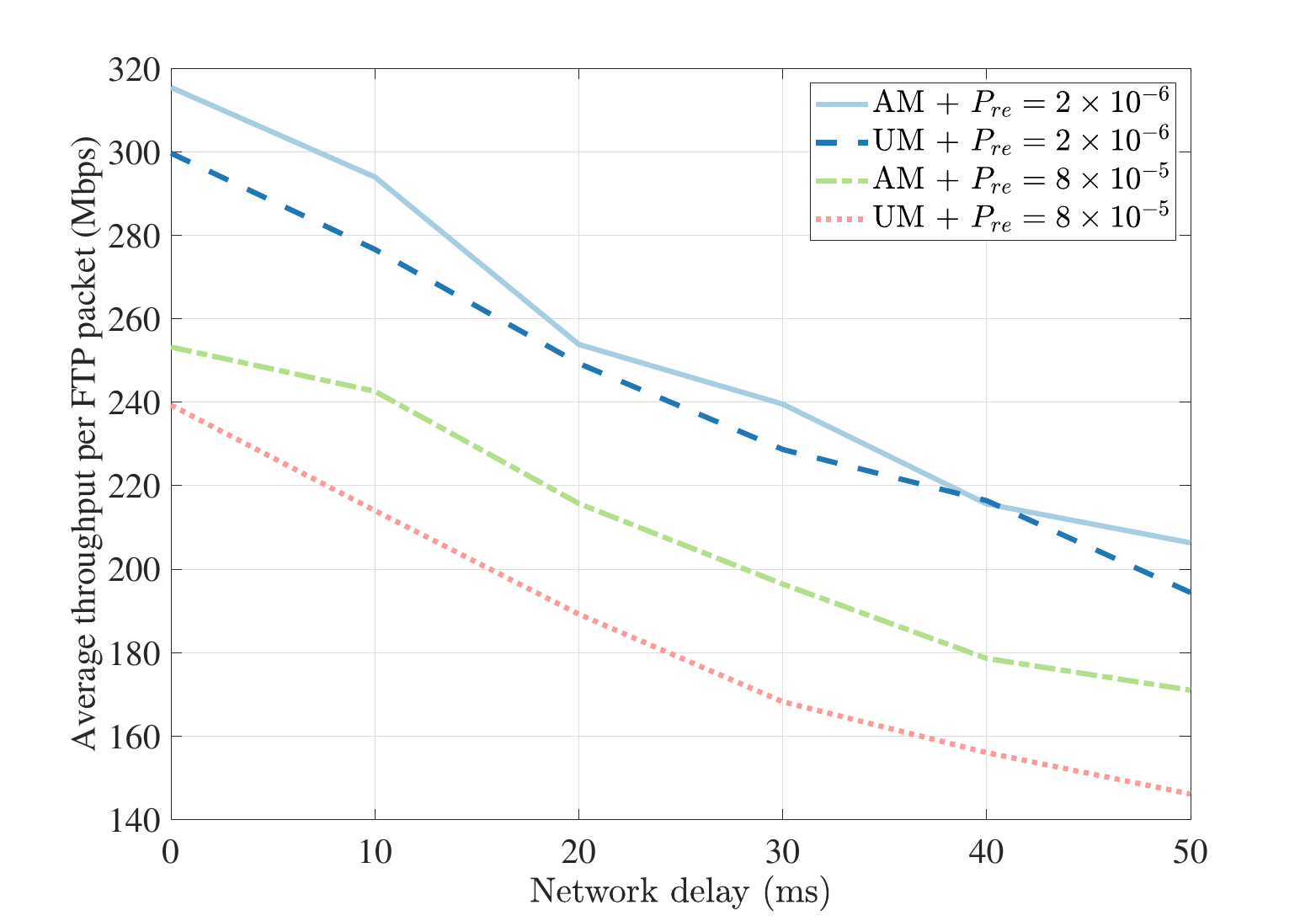}
\caption{Average user throughput (Mbps) versus Network delay (ms)}
\label{fig_wh}
\end{figure}

For the plots in Figures \ref{fig_sim1}  and \ref{fig_sim3}, the simulations were run with a fixed network delay of \SI{10}{ms}. The network delay is the delay between the radio access network and the application server. This network delay includes transport, core network processing, and Internet delays towards the TCP server. In Fig.~\ref{fig_wh}, the behavior of user throughput for a sweep of network delay when the TCP CUBIC congestion control algorithm is used is presented. 

There is a pattern or trend of reduction in user throughput as the network delay is increased from \num{0} to \num{50} ms. The effect of residual error rate is also seen in the performance gap between the $P_{re} = 2\times 10^{-6}$  and $P_{re} =8\times 10^{-5}$ curves. The slight gain of having a recovery loop at the RLC layer when the residual error is relatively higher ($P_{re} =8\times 10^{-5}$) is also seen in the performance gap between $UM +$ $P_{re} = 8\times 10^{-5}$ and $AM +$ $P_{re} = 8\times 10^{-5}$ curves. Regardless of network delay, the relative gain of using RLC AM is very limited as long as the residual error rate is kept lower than $8\times 10^{-5}$ or lower.

\section{Recommendations towards 6G}
As shown HARQ is the foundation for low-latency and spectrally efficient data recovery for dynamic radio environments. HARQ is fundamental and needed when using fast dynamic link adaptation.
In the previous, we investigated what residual error rate is needed for various levels of throughput and delay degradation with and without a higher layer data recovery mechanisms on top of HARQ. As a good design target, the combined effects of wrongly interpreting a failed transmission as a correctly received one, i.e. the residual error rate, should be bounded by $P_{re}<5\times 10^{-5}$.

\begin{figure}[!t]
\centering
\includegraphics[width=\linewidth,trim={0 0 0 0cm},clip]{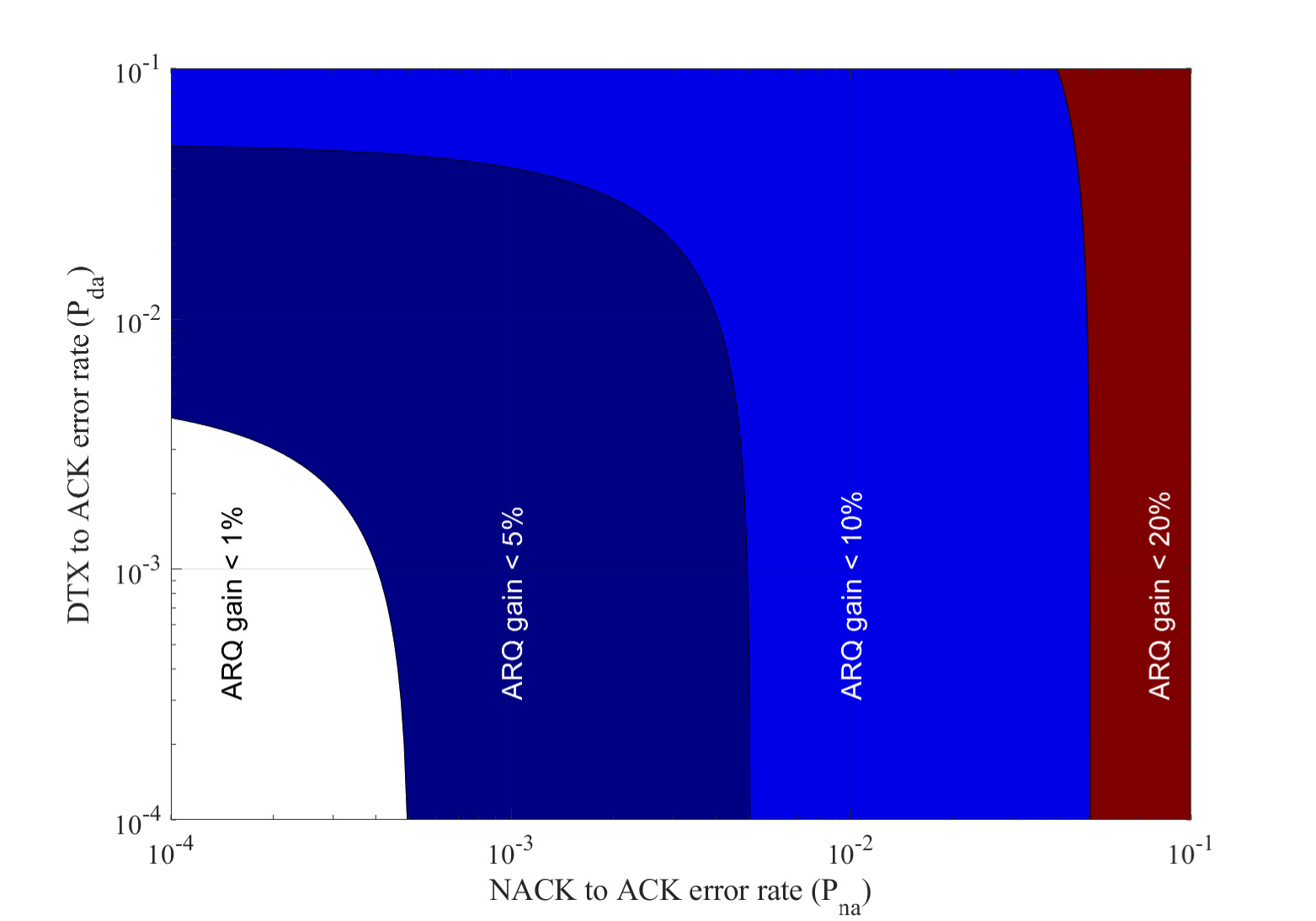}
\caption{Relation between $P_{na}$ and $P_{da}$ to observed throughput degradation without ARQ to supplement HARQ}
\label{fig_countour}
\end{figure}

When thinking about the design for 6G, we can look at these results as a function of how well we need to design the 6G control channels using the relations captured by \eqref{eq_1}.  Assuming typical values for reliability for the downlink control channel that will carry the dynamic scheduling grant, $P_{ch}=1\%$, as well as a typical operational BLER, $P_i=10\%$ for good tradeoff among spectral efficiency, delays, and peak rates, we can derive suitable targets for the design for $P_{na}$ and $P_{da}$.  Results are shown in Fig.~\ref{fig_countour}. As one example, to operate without ARQ (e.g. to avoid ARQ-related latency increase) and minimize the effective user throughput loss with TCP to $<5\%$ we can see the boundaries for $P_{da}$ and $P_{na}$ design in the white area. 

In terms of the typical performance numbers as also reported in \cite{7063628}, the design of NACK to ACK error $<5\times 10^{-3}$ is a feasible task. The main design challenge is mainly around DTX to ACK errors, but it should be noted that the effect of these can also be improved by making the downlink control channel more robust, e.g. work towards $P_{ch}<1\%$. Other schemes for implicit or explicit signaling and handling the NACK and ACK process should also be considered.

Higher layer data recovery will still be needed in cases where low residual errors cannot be guaranteed to yield significant performance impact for Internet users leveraging TCP. With increased expectations of low-latency support in the \SI{10}{ms} range, alternatives to RLC ARQ need to be considered, including methods for packet-level coding as considered in \cite{8554264}.

\section{Conclusion}
In this paper, we have shown that HARQ, despite its inherent complexity, remains a key element for data recovery when designing 6G networks, and its absence would lead to significant spectral efficiency and coverage performance losses. Practical HARQ designs will result in a residual error, and we have shown that keeping this rate $<\num{5e-5}$, avoids significant losses to TCP throughput regardless of the network delays and the TCP protocol used. From this, we have derived suitable design targets for 6G control channel design and methodologies.  

\section*{Acknowledgment}
This work has been partly funded by the European Union’s Horizon Europe research and innovation programme under grant agreement No. 101095759 (Hexa-X-II).

\ifCLASSOPTIONcaptionsoff
  \newpage
\fi

\bibliographystyle{IEEEtran}

\end{document}